# Dynamics of Gender Bias within Computer Science

*Thomas J. Misa*

Misa directed the University of Minnesota's Charles Babbage Institute (2006-17) and was President of the Society for the History of Technology (2019-20). His most recent book is *Leonardo to the Internet* (Johns Hopkins University Press, 2022, third ed.).

**Abstract**
A new dataset (N = 7,456) analyzes women's research authorship in the Association for Computing Machinery's founding 13 Special Interest Groups or SIGs, a proxy for computer science. ACM SIGs expanded during 1970-2000; each experienced increasing women's authorship. But diversity abounds. Several SIGs had fewer than 10% women authors while SIGUCCS (university computing centers) exceeded 40%. Three SIGs experienced accelerating growth in women's authorship; most, including composite ACM, had decelerating growth. This research may encourage reform efforts, often focusing on general education or workforce factors (across "computer science"), to examine under-studied dynamics within computer science that shaped changes in women's participation.

There is some good news concerning gender bias in computer science. Women's participation in computer science has improved in the years since the low point in 2009, when the respected Computing Research Association (CRA) Taulbee annual survey of PhD-granting departments in North America found women gaining just 11.3 percent of undergraduate computer science degrees, with comparable NSF data also registering a longitudinal low point that year.[1] Sustained reform efforts by policy actors, professional organizations, and thousands of practitioners have addressed the male-heavy slant of computer science, with notable success stories at Harvey Mudd, Carnegie Mellon, University of Washington, University of California–Berkeley, and other colleges and universities, as well as the landmark annual Grace Hopper Celebration of Women in

---

[1] The CRA Taulbee Survey for 2008-2009 (table 9a on page 11) tabulates North American PhD-granting departments' undergraduate degrees awarded in computer science (11.3% women), computer engineering (8.7%), and information (13.1%); see Computing Research Association–Taulbee Survey (2001 et seq.) at cra.org/resources/taulbee-survey/ (Oct. 2023). NSF data in 2009 registered a low point of women gaining 17.7% of BS computer science degrees.

Computing.[2] The Association for Computing Machinery (ACM) itself, the Alfred P. Sloan Foundation, the CRA's Committee on Women in Computing Research (CRA-W), Anita Borg Institute, National Center for Women and Information Technology, National Science Foundation, American Association for University Women, and numerous others have confronted the pernicious gender bias that seems endemic in computer science.[3] Compared with the 2009 nadir, North American women are receiving modestly greater proportions of bachelor's, master's, and doctoral computer-science degrees.[4] And we now have rich international data that helps round

---

[2] See Jane Margolis and Allan Fisher, *Unlocking the Clubhouse: Women in Computing* (Cambridge: MIT Press, 2001); Christine Alvarado, Zachary Dodds, and Ran Libeskind-Hadas, "Increasing Women's Participation in Computing at Harvey Mudd College," *ACM Inroads* 3 no. 4 (December 2012): 55–64 at doi.org/10.1145/2381083.2381100 ; Carol Frieze and Jeria L. Quesenberry, "How Computer Science at CMU is Attracting and Retaining Women," *Communications of the ACM* 62 no. 2 (February 2019): 23-26 at doi.org/10.1145/3300226 ; Sarah McBride, "Glimmers of Hope for Women in the Male-Dominated Tech Industry," *Bloomberg Technology* (March 8, 2018) at https://www.bloomberg.com/news/articles/2018-03-08/glimmers-of-hope-for-women-in-the-male-dominated-tech-industry (Feb. 2023).

[3] See William Aspray, *Participation in Computing: The National Science Foundation's Expansionary Programs* (Basel: Springer, 2016) at doi.org/10.1007/978-3-319-24832-5 ; and William Aspray, *Women and Underrepresented Minorities in Computing: A Historical and Social Study* (Basel: Springer, 2016) at doi.org/10.1007/978-3-319-24811-0 ; Jeff Forbes, Allyson Kennedy, Margaret Martonosi, and Fernanda Pembleton, "Expanding the Pipeline: Roadmap of CISE's Efforts to Broaden Participation in Computing Through the Years," *Computing Research News* 35 no. 2 (February 2023) at cra.org/crn/2023/02/expanding-the-pipeline-roadmap-of-cises-efforts-to-broaden-participation-in-computing-through-the-years/ (Feb. 2023).

[4] According to the most recent CRA Taulbee survey (2022), women received 22.7% of bachelor's, 30.9% of master's, and 22.9% of doctoral degrees (non-binary or other genders were reported as 0-0.2%). By contrast, in 2008-09, women received 11.1% of bachelor's, 27.4% of master's, and 21.2% of doctoral degrees. The data for 2022 sums computer science, computer engineering, and information degrees. See data at cra.org/resources/taulbee-survey/ (January 2024).

out the US focus that has dominated research so far.[5]

All the same, there is a long way to go until computer science regains the peak of women's participation that occurred in the mid-1980s, when women gained 37% of US baccalaureate degrees in computer science and constituted 38% of the US white-collar information-technology workforce.[6] The recent 2022 CRA–Taulbee study indicates that women collected 22.9% of North American undergraduate computer science degrees; the NSF found this figure back in *1977*.

---

[5] Vashti Galpin, "Women in Computing around the World," *SIGCSE Bulletin* 34 no. 2 (June 2002): 94-100 at doi.org/10.1145/543812.543839 ; Joel C. Adams, Vimala Bauer, and Shakuntala Baichoo, "An Expanding Pipeline: Gender in Mauritius," Proceedings of the 34th SIGCSE Technical Symposium on Computer Science Education (SIGCSE '03): 59-63 at doi.org/10.1145/611892.611932 ; Hasmik Gharibyan and Stephan Gunsaulus, "Gender Gap in Computer Science does not exist in one former Soviet republic [Republic of Armenia]: results of a study," Proceedings of the 11th annual SIGCSE Conference on Innovation and Technology in Computer Science Education (ITICSE '06): 222-226 at doi.org/10.1145/1140124.1140184 ; J. McGrath Cohoon and William Aspray, ed., *Women and Information Technology: Research on Underrepre*sentation (Cambridge: MIT Press, 2006), 183-203; Vivian Anette Lagesen, "A Cyberfeminist Utopia?: Perceptions of Gender and Computer Science among Malaysian Women Computer Science Students and Faculty," *Science, Technology, & Human Values* 33 no. 1 (2008): 5-27 at doi.org/10.1177/0162243907306192 ; Ulf Mellström, "The Intersection of Gender, Race and Cultural Boundaries, or Why is Computer Science in Malaysia Dominated by Women?" *Social Studies of Science* 39 no. 6 (2009): 885–907 at doi.org/10.1177/0306312709334636 ; Ksenia Tatarchenko, "'The Computer Does Not Believe in Tears': Soviet Programming, Professionalization, and the Gendering of Authority," *Kritika: Explorations in Russian and Eurasian History* 18 no. 4 (2017): 709-739 at doi.org/10.1353/kri.2017.0048 ; Moreno Marzolla and Raffaela Mirandola, "Gender Balance in Computer Science and Engineering in Italian Universities," Proceedings of the 13th European Conference on Software Architecture - Volume 2 (ECSA '19): 82-87 at doi.org/10.1145/3344948.3344966 ; Hyomin Kim, Youngju Cho, Sungeun Kim, and Hye-Suk Kim, "Women and Men in Computer Science: Geeky Proclivities, College Rank, and Gender in Korea," *East Asian Science, Technology and Society* 12 no. 1 (2018): 33-56 at doi.org/10.1215/18752160-4206046; Alexander Repenning, Anna Lamprou, Serge Petralito, and Ashok Basawapatna, "Making Computer Science Education Mandatory: Exploring a Demographic Shift in Switzerland," Proceedings of the 2019 ACM Conference on Innovation and Technology in Computer Science Education (ITiCSE '19): 422-428 at doi.org/10.1145/3304221.3319758 ; Maria Kordaki and Ioannis Berdousis, "Gender Differences in [Greek] Computer Science Departments," in Arthur Tatnall, ed., *Encyclopedia of Education and Information Technologies* (Cham: Springer, 2020), at doi.org/10.1007/978-3-030-10576-1_184 ; Bidisha Chaudhuri, Meenakshi D'Souza, and Janaki Srinivasan, "Bringing the Missing Women Back: CS Education for Women in India's Engineering Institutions," *Communications of the ACM* 65 no. 11 (November 2022): 65-67 at doi.org/10.1145/3550490.

[6] Caroline Clarke Hayes, "Computer Science: The Incredible Shrinking Woman," in Thomas J. Misa, ed., *Gender Codes: Why Women are Leaving Computing* (Hoboken, NJ: Wiley, 2010), 25-49 at doi.org/10.1002/9780470619926.ch2 . Workforce data from the Bureau of Labor Statistics needs careful inspection; for example, one industry survey appeared to find a significant *decline* in women's share of the IT workforce from 41% to 34.9% (1996–2002). Yet, "in 1996, the higher percentage was due largely to greater numbers of women in the administrative IT positions of Data Entry Keyers [81.8% women] and Computer Operators [46.8% women]. Typically, these jobs require less formal education and experience, and command demonstrably lower pay than other positions listed in this BLS category." Removing these two lower-pay positions resulted in *stable* women's share of the IT workforce, namely 25% and 25.3% (1996–2002); see ITAA, *Report of the ITAA Blue Ribbon Panel on IT Diversity* (Arlington, VA: National IT Workforce Convocation, 2003), 11 archived at https://web.archive.org/web/20230228192854/https://www.ictal.org/public/downloads-old/2002-8/ITAA_Diversity_Report_2003.pdf .

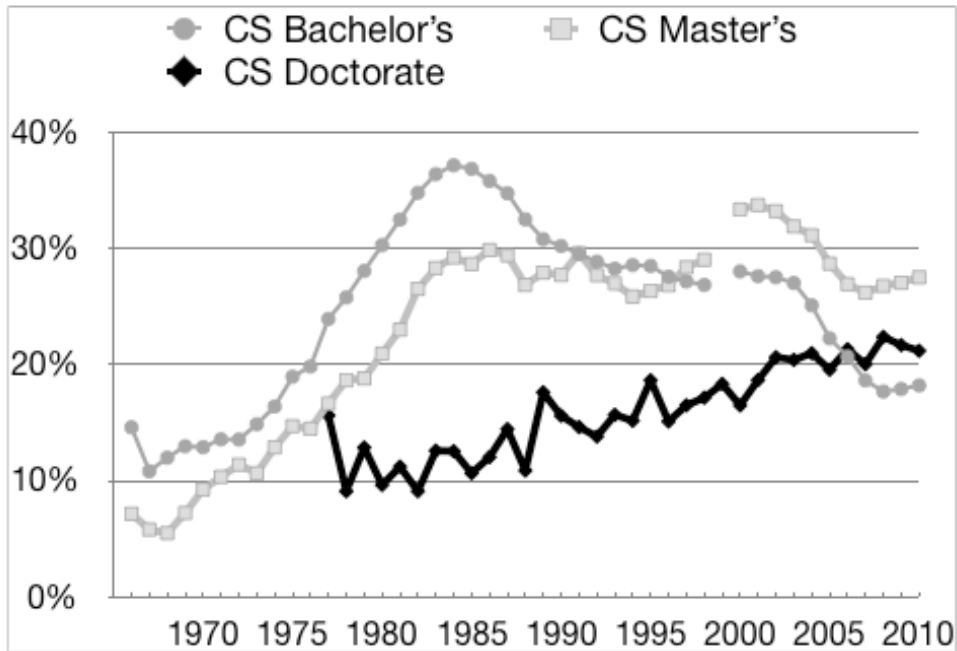

Figure 1: Women's share of Computer Science degrees (1966-2010)

Source: National Science Foundation and National Center for Science and Engineering Statistics, *Science and Engineering Degrees: 1966–2010: Detailed Statistical Tables NSF 13-327* (Arlington: NSF, 2013), table 33, at www.nsf.gov/statistics/nsf13327/. NSF data includes one female doctoral degree in 1973. The table's note (a) states "In the Survey of Earned Doctorates (SED), data on computer sciences were not collected separately from mathematics until 1978, and complete data on computer sciences are not available from the SED until 1979. Data shown for 1966–78 are from the Integrated Postsecondary Education Data System Completions Survey."

There is wide agreement that persisting gender bias in computer science is a problem—resulting in poorly designed computer systems, entrenching unhealthy power differentials, unduly restricting women's positive influences on computing, and reducing the size and diversity of the computing workforce. Of course, there are many pathways into the computing workforce, including formal and informal education; computer science, while influential in defining research frontiers and shaping future developments, is just one route.[7] All the same, computer science as an academic field has been and remains incredibly diverse in subject matter; the gender composition of individual subfields within computer science has been understudied. Specifically, comparative data on women's research contributions to different subfields is sorely needed. This article presents a new dataset based on analysis of one of the two leading professional organizations in computing, the Association for Computing Machinery, to measure this diversity in women's participation in computer science as research authors. Implications for gender reform efforts are outlined in the conclusion.

**Gender bias in computing**

Social science and gender-studies scholars have offered insight into numerous mechanisms and processes that lead to gender bias in computing. These include gender-slanted stereotypes and societal models; expectations about gender roles, including occupational stereotyping; masculine or "bro" computer culture, including overt discrimination, marginalization, and harassment; women's "hidden" authorship, where women's contributions may be under-valued or under-reported by men claiming authorship; organizational gender bias, including differentials in hiring, promotion, and salary; weak employee representation entities (such as ineffective Human

---

[7] As Freeman and Aspray put it, "most IT workers receive their formal education in fields other than computer science"; Peter Freeman and William Aspray, *The Supply of Information Technology Workers in the United States* (Computing Research Association, 1999), quote p. 17 at archive.cra.org/reports/wits/it_worker_shortage_book.pdf (Feb. 2023).

Resources departments); and several others that have been studied in some detail.[8] The current fascination for "coding" boot camps, non-traditional programming classes, and even hackathons as a potential route for women (and other 'disadvantaged' groups) to directly take up computing careers has recently received much-needed critical assessment. Kate Miltner in *Information & Culture* points out deep comparisons between the present-day programming craze and 1960s-era electronic data programming (EDP) schools where, as today, "programming was often positioned as a 'fix' for labor exclusion and a gateway to social mobility for a variety of minoritized groups, including women, Black people, the disabled, the incarcerated, and the economically marginalized."[9] Recently, attention to "algorithmic bias" suggests the likelihood of gender, race, class, and other pernicious biases being even further entrenched in computing systems and the wider society with significant influence but little transparency.[10] For many, a widely circulated meme from machine-learning research "Man is to computer programmer as woman is to homemaker" indicates that something is seriously wrong.[11]

---

[8] Paula Mählck, "Mapping Gender Differences in Scientific Careers in Social and Bibliometric Space," *Science, Technology, & Human Values* 26 no. 2 (2001): 167-190 at doi.org/10.1177/016224390102600203 ; Myria Watkins Allen, Deborah J. Armstrong, Cynthia K. Riemenschneider, and Margaret F. Reid, "Making Sense of the Barriers Women Face in the Information Technology Work Force: Standpoint Theory, Self-disclosure, and Causal Maps," *Sex Roles* 54 (June 2006), 831-844 at doi.org/10.1007/s11199-006-9049-4 ; Francesca Bray, "Gender and Technology," *Annual Review of Anthropology* 36 (2007):1-21 at doi.org/10.1146/annurev.anthro.36.081406.094328 ; Janet Abbate, *Recoding Gender: Women's Changing Participation in Computing* (Cambridge: MIT Press, 2012); Mar Hicks, *Programmed Inequality: How Britain Discarded Women Technologists and Lost Its Edge in Computing* (Cambridge: MIT Press, 2017); Sharla Alegria, "Escalator or Step Stool? Gendered Labor and Token Processes in Tech Work," *Gender and Society* 33 no. 5 (October 2019): 722-745 at doi.org/10.1177/0891243219835737 ; Thomas S. Mullaney, Benjamin Peters, Mar Hicks, and Kavita Philip, eds., *Your Computer Is on Fire* (Cambridge: MIT Press, 2021); Emerson Murphy-Hill, Ciera Jaspan, Carolyn Egelman, and Lan Cheng, "The Pushback Effects of Race, Ethnicity, Gender, and Age in Code Review," *Communications of the ACM* 65 no. 3 (2022): 52-57 at doi.org/10.1145/3474097 ; Donna Bridges, Elizabeth Wulff, and Larissa Bamberry, "Resilience for Gender Inclusion: Developing a Model for Women in Male-Dominated Occupations," *Gender, Work, and Organization* 30 (2023): 263-279 at doi.org/10.1111/gwao.12672.

[9] R. Arvid Nelsen, "Concern for the 'Disadvantaged': ACM's Role in Training and Education for Communities of Color (1958–1975)," in Thomas J. Misa, ed., *Communities of Computing: Computer Science and Society in the ACM* (New York: ACM Books, 2016), 229-58 at doi.org/10.1145/2973856.2973867 ; Kate M. Miltner, "Girls Who Coded: Gender in Twentieth Century U.K. and U.S. Computing," *Science, Technology, & Human Values* 44 no. 1 (2018): 161-176 at doi.org/10.1177/0162243918770287 ; Janet Abbate, "Code Switch: Alternative Visions of Computer Expertise as Empowerment from the 1960s to the 2010s," *Technology and Culture* 59 no. 4 supplement (October 2018): S134-S159 at doi.org/10.1353/tech.2018.0152; Kate M. Miltner, "Everything Old Is New Again: A Comparison of Midcentury American EDP Schools and Contemporary Coding Bootcamps," *Information & Culture* 57 no. 3 (2022): 255-282, quote 256, at doi.org/10.7560/IC57302 .

Reform efforts inspired by gender studies scholarship have offered two general areas for intervention. On the societal level, reformers have directed critical attention to educational initiatives (such as programming classes and coding camps, noted above), investigation of discriminatory recruitment and hiring practices, and constructive confrontation of societal images (such as white male 'hackers') and other unwelcome biases.[12] Professional and academic efforts have typically focused on reform of "computer science." Numerous researchers have investigated the gender dynamics of "computer science as a field in general" (as one study puts it) often at specific universities or in different national contexts.[13] At Carnegie Mellon, one of the success stories, the reform efforts drew "close attention to culture and environment" through "taking a cultural approach rather than a gender difference approach" that included dropping programming as an undergraduate entry requirement for computer science, sustaining long-term positive academic leadership (1999-present), hiring female faculty with specific experience and expertise in women's advocacy, and supporting an active women's student group in computer science. Roughly two decades of scholarship combining gender theory, empirical research, and attentiveness to educational, professional, and workplace contexts has grounded institutional reform efforts in recent years. Beginning in 2017, the National Science Foundation has required specific broadening participation in computing activities in all research proposals to seven programs in the core Computer and Information Science and Engineering (CISE) directorate.[14]

---

[10] Cathy O'Neil, *Weapons of Math Destruction: How Big Data Increases Inequality and Threatens Democracy* (New York: Crown, 2016); Safiya Umoja Noble, *Algorithms of Oppression: How Search Engines Reinforce Racism* (New York: New York University Press, 2018); Meredith Broussard, *Artificial Unintelligence: How Computers Misunderstand the World* (Cambridge: MIT Press, 2018); Susan Leavy, "Gender Bias in Artificial Intelligence: The Need for Diversity and Gender Theory in Machine Learning," 2018 ACM/IEEE 1st International Workshop on Gender Equality in Software Engineering (2018): 14-16 at doi.org/10.1145/3195570.3195580; Ruha Benjamin, *Race After Technology: Abolitionist Tools for the New Jim Code* (2019); Thomas S. Mullaney, Benjamin Peters, Mar Hicks, and Kavita Philip, eds., *Your Computer Is on Fire* (Cambridge: MIT Press, 2021); Juan De Lara, "Race, Algorithms, and the Work of Border Enforcement," *Information & Culture* 57 no. 2 (2022): 150-168 at muse.jhu.edu/article/856982; Janet Abbate and Stephanie Dick, eds., *Abstractions and Embodiments: New Histories of Computing and Society* (Baltimore: Johns Hopkins University Press, 2022), esp. chapters 5 and 17.

[11] Tolga Bolukbasi, Kai-Wei Chang, James Zou, Venkatesh Saligrama, Adam Kalai, "Man is to Computer Programmer as Woman is to Homemaker? De-biasing Word Embeddings," 2016 at doi.org/10.48550/arXiv.1607.06520 ; Marzieh Babaeianjelodar, Stephen Lorenz, Josh Gordon, Jeanna Matthews, and Evan Freitag, "Quantifying Gender Bias in Different Corpora," Companion Proceedings of the Web Conference 2020 (WWW '20): 752-759 at doi.org/10.1145/3366424.3383559 .

[12] Overviews are Vivian Anette Lagesen, "The Strength of Numbers: Strategies to Include Women into Computer Science," *Social Studies of Science* 37, no. 1 (Feb., 2007): 67-92 at doi.org/10.1177/0306312706063788 ; Els Rommes, Corinna Bath and Susanne Maass "Methods for Intervention: Gender Analysis and Feminist Design of ICT," *Science, Technology, & Human Values* 37, no. 6 (2012): 653-662 at doi.org/10.1177/0162243912450343 ; a case study is Judith Stepan-Norris and Jasmine Kerrissey, "Enhancing Gender Equity in Academia: Lessons from the ADVANCE Program," *Sociological Perspectives* 59, no. 2 (2016): 225-245 at doi.org/10.1177/0731121415582103 .

[13] Maureen Biggers, Anne Brauer, and Tuba Yilmaz, "Student Perceptions of Computer Science: A Retention Study Comparing Graduating Seniors with CS Leavers," *Proceedings of the 39th SIGCSE Technical Symposium on Computer Science Education* (SIGCSE '08) (March 2008) 402-406 at doi.org/10.1145/1352135.1352274 ; Antonio M. Lopez Jr., Kun Zhang, and Frederick G. Lopez, "Cultural Representations of Gender Among U. S. Computer Science Undergraduates: Statistical and Data Mining Results," *Proceedings of the 39th SIGCSE Technical Symposium on Computer Science Education* (SIGCSE '08) (March 2008), pp 407–411 at doi.org/10.1145/1352135.1352275 .

      A less studied dimension of diversity in computing is the changing dynamics of gender bias, for it is not a static phenomenon. To cite one familiar example, since data became first available in the mid-1960s, women in the US—at different points in time—have collected quite different proportions of undergraduate computer science degrees: 13% in the mid 1960s, 37% in the mid-1980s, below 18% in 2008-9, and slowly *climbing* to just over 20% in recent years (compare Figure 1). Appeals to a static conception of gender bias, most everywhere the same and virtually unchanging, cannot properly recognize these significant changes across time. A dynamic conception of gender bias in computing is not only good history; it may suggest possible paths for future change. After all, reform efforts need greater insight into processes of change in addition to awareness of the persistent oppression of static structures; gender scholars have paid more attention to the structures of oppression and less attention to changes across time.

---

[14] Carol Frieze, Jeria L. Quesenberry, Elizabeth Kemp, and Anthony Velázquez, "Diversity or Difference? New Research Supports the Case for a Cultural Perspective on Women in Computing," *Journal of Science Education and Technology* 21 no. 4 (2012): 423-439 at doi.org/10.1007/s10956-011-9335-y ; Carol Frieze and Jeria L. Quesenberry, "How Computer Science at CMU Is Attracting and Retaining Women," *Communications of the ACM* 62 no. 2 (February 2019): 23-26, quote 23 at doi.org/10.1145/3300226; Tracy Camp, et al., "The New NSF Requirement for Broadening Participation in Computing (BPC) Plans: Community Advice and Resources," in *Proceedings of the 50th ACM Technical Symposium on Computer Science Education* (SIGCSE'19) Feb. 27-March 2, New York: ACM, 2019, pages 332-333 at doi.org/10.1145/3287324.3287332; policy documents are "CISE Strategic Plan for Broadening Participation" (18 November 2012) at www.nsf.gov/cise/oad/cise_bp.jsp and "Broadening Participation in Computing (BPC)" (15 October 2022) at www.nsf.gov/cise/bpc/. The CISE Strategic Plan specifically cites such gender research findings, approaches and concepts as implicit bias, communities of practice, and institutional transformations.

A related question is the differential effects of general "environmental" factors (such as education and workforce trends, social movements like the 1970s women's movement, and broader cultural trends) compared with forces and trends "within" the field of computing. A recent quantitative study found that nearly two-thirds of gender segregation in the college-educated U.S. workforce may be traced to patterns of gender segregation *within* the fields of study with the remaining one-third traced to between-field effects.[15]

Gender theory might shed some conceptual light on computer science: both fields are active areas of research and neither are unitary. "Feminism is plural; there are many feminisms, and they differ in their positive visions, methodologies, collective ends, and situated concerns," declares a manifesto published in *Debates in the Digital Humanities* (2023);[16] a similar observation might serve to bridge the stunningly diverse subfields in computing, ranging from near-hardware level studies of computer architecture to the heady abstractions of computer-science theory. This present article offers a new approach to assess diversity within computer science as a research field.

"Computer science" frequently appears as an object of reformers addressing underrepresentation in the field, as well as introductory textbook authors and professional-society discourse on ethics, whereas it is rare to assess diversities across the burgeoning field.

---

[15] Haowen Zheng, Kim A. Weeden, "How Gender Segregation in Higher Education Contributes to Gender Segregation in the U.S. Labor Market," *Demography* 60, no. 3 (June 2023): 761-784 at doi.org/10.1215/00703370-10653728; the authors define within-field segregation as "gender differences in occupational destinations of those who graduate with degrees in the same field" (p. 762) while their finding implies that "integrating higher education (e.g., by increasing women's representation in STEM majors) will reduce but not eliminate gender segregation in labor markets" (pp. 761 and 780)

[16] Tonia Sutherland, Marika Cifor, T. L. Cowan, Jas Rault, and Patricia Garcia, "The Feminist Data Manifest-NO: An Introduction and Four Reflections," in Matthew K. Gold and Lauren F. Klein, eds., *Debates in the Digital Humanities 2023* (Minneapolis: University of Minnesota Press, 2023), 120-39 quote p. 120. Some but not all computer scientists would embrace the manifesto's "refusal of . . . an inheritance of 'imperialist white-supremacist capitalist patriarchy'."

One recent large-scale study addresses such differences *within* computer science.[17] Nicholas Laberge and coauthors investigate the correlations or intersectionality of factors such as gender, race, class, and subfield prestige.[18] Tracing nearly 7,000 computer-science faculty (2010-2018), the authors find significant and systematic differences between high-prestige subfields like Theory of Computer Science or Programming Languages compared with lower-prestige ones like Interdisciplinary Computing or Human-Computer Interaction in their proportion of women faculty (and other measures). Such intersectionality may reproduce gender and other biases: "most searches in computing remain subfield-specific" (despite evidence that non-field-specific searches can result in greater diversity in the candidate pool), "faculty searches in subfields with fewer women than other subfields are less likely to increase a department's gender diversity" (p. 47). Put simply, faculty searches in specific subfields—especially male-dominated high-prestige subfields—tend to reproduce gender bias.

---

[17] Nicholas Laberge, K. Hunter Wapman, Allison C. Morgan, Sam Zhang, Daniel B. Larremore, and Aaron Clauset, "Subfield Prestige and Gender Inequality among U.S. Computing Faculty," *Communications of the ACM* 65, no. 12 (December 2022): 46-55 at doi.org/10.1145/3535510 . On *subfields* constituting computer science; see Brent K. Jesiek, "The Origins and Early History of Computer Engineering in the United States," *IEEE Annals of the History of Computing* 35 no. 3 (2013): 6-18 at doi.org/10.1109/MAHC.2013.2 ; David Nofre, Mark Priestley, and Gerard Alberts, "When Technology Became Language: The Origins of the Linguistic Conception of Computer Programming, 1950-1960," *Technology and Culture* 55, no. 1 (2014): 40-75 at doi.org/10.1353/tech.2014.0031; Elizabeth R. Petrick, "A Historiography of Human–Computer Interaction," *IEEE Annals of the History of Computing* 42 no. 4 (2020): 8-23 at doi.org/10.1109/MAHC.2020.3009080 ; Liesbeth De Mol, "Logic, Programming, and Computer Science: Local Perspectives," *IEEE Annals of the History of Computing* 43 no. 4 (2021): 5-9 at doi.org/10.1109/MAHC.2021.3121578 ; and David Nofre "'Content Is Meaningless, and Structure Is All-Important': Defining the Nature of Computer Science in the Age of High Modernism, c. 1950–c. 1965," *IEEE Annals of the History of Computing* 45 no. 2 (2023): 29-42 at doi.org/10.1109/MAHC.2023.3266359 .

[18] Intersectionality, an active topic in the social sciences, is an emerging area in computer science: Eileen M. Trauth, Curtis C. Cain, K.D. Joshi, Lynette Kvasny, and Kayla M. Booth, "The Influence of Gender-Ethnic Intersectionality on Gender Stereotypes about IT Skills and Knowledge," *SIGMIS Database* 47, no. 3 (August 2016): 9-39 at doi.org/10.1145/2980783.2980785 ; Ari Schlesinger, W. Keith Edwards, and Rebecca E. Grinter, "Intersectional HCI: Engaging Identity through Gender, Race, and Class," Proceedings of the 2017 CHI Conference on Human Factors in Computing Systems (CHI '17): 5412-27 at doi.org/10.1145/3025453.3025766 ; Joy Buolamwini and Timnit Gebru, "Gender Shades: Intersectional Accuracy Disparities in Commercial Gender Classification," *Proceedings of Machine Learning Research* 81 (2018):1-15 at https://proceedings.mlr.press/v81/buolamwini18a.html (Feb. 2023) and archived at https://web.archive.org/web/20240111214231/https://proceedings.mlr.press/v81/buolamwini18a.html; Emerson Murphy-Hill, Ciera Jaspan, Carolyn Egelman, and Lan Cheng, "The Pushback Effects of Race, Ethnicity, Gender, and Age in Code Review," *Communications of the ACM* 65, no. 3 (March 2022): 52-57 at doi.org/10.1145/3474097 .

The data presented in this present article also examines subfield differences within computer science. But while Laberge et al. use algorithmically created categories—their "subfields" are created by their machine-learning algorithms working across the set of computer-science articles—this dataset instead uses "natural" categories that were created by the ACM SIG's themselves and sustained by them across the study years. My approach avoids the possible circularities in using algorithmically created categories both to identify the subfields and to assess their prestige, and so possibly identifying artificial correlations between artificial constructs. Unfortunately, Laberge et al.'s use of just 8 subfields to describe computer science (theory of computer science, programming languages, numerical and scientific computing, systems, computational learning, software engineering, interdisciplinary computing, and human-computer interaction) simply does not align with the *19* subfields used in the long-running Taulbee surveys of North American Ph.D. granting CS departments.[19] Moreover, it is somewhat unclear why HCI, an active area of high-impact research with stringent conference acceptance rates (a common measure of subfield prestige in CS), should score "low" in prestige by Laberge et al.; while Programming Languages, an established and mature subfield, should score "high." The 13 ACM SIGs examined in this article have close alignment with the Taulbee categories.[20]

**New Data on Women in ACM SIGs**

To better understand women's changing participation in computer science, this article assesses women's changing authorship in the research literature of ACM's founding 13 Special Interest Groups from 1970 to 2000. It analytically and graphically demonstrates dramatic variation between different subfields of computer science, and points to significantly differing receptivity to women as members of the research community. One implication is that gender-reform efforts focusing on the whole of "computer science" may miss the mark. Right at hand, there are instances of some ACM SIGs' striking openness to women's research-article authorship, whereas other SIGs have been less so. This disaggregated data, based on "natural" ACM-created categories of computer science, may also serve to suggest hypotheses about the influence of external (environmental) or internal (subfield-specific) dynamics that shape women's research authorship and their participation in the computer science research community. The data provides longitudinal and comparative insight on the dynamics of gender bias in the research literature of computer science.

---

[19] See Betsy Bizot, "Expanding the Pipeline: Gender and Ethnic Differences in PhD Specialty Areas," *CRA Reports* 31 no. 7 (August 2019) at cra.org/crn/2019/08/expanding-the-pipeline-gender-and-ethnic-differences-in-phd-specialty-areas/ (Feb. 2023). The two datasets have "point" overlaps (labeling subfields with the same names) but they cannot be systematically compared.

[20] The Taulbee survey's subfield categories Computing Education, Databases / Information Retrieval, Graphics / Visualization, Networks, and Operating Systems are explicitly included in the 13 original ACM SIGs (ACM SIGCOMM was founded for data communication and networking)—while these subfields do *not* appear at all in Laberge et al., which instead offers expansive generic subfields such as "Systems" and "Interdisciplinary Computing" (respectively, fully 27% and 16% of their entire dataset). An alternative ranking of nine CS subfields is examined in Yifan Qian, Wenge Rong, Nan Jiang, Jie Tang, and Zhang Xiong, "Citation Regression Analysis of Computer Science Publications in Different Ranking Categories and Subfields," *Scientometrics* 110 (2017): 1351-1374 at doi.org/10.1007/s11192-016-2235-4.

The Association for Computing Machinery was founded in 1947 and quickly became one of two leading professional and scientific organizations in digital computing active worldwide.[21] The other, emerging from the Institute of Radio Engineers' Professional Group on Electronic Computers, eventually became the IEEE's Computer Society in 1971.[22] Today, the two societies are friendly competitors in international computing research, education, and professional activities. During the 1960s ACM created more than a dozen special interest groups, or SIGs, sometimes initially as special interest committees or SICs. In 1960 ACM was largely a unitary organization; by 1970 and continuing to this day SICs and SIGs are active arms of the ACM in sponsoring research conferences, building membership, and publishing research literature. SICs and SIGs, suggested ACM President Jean Sammet, herself a major figure in the history and methodology of programming languages, "provide the advantage of what are inherently smaller technical societies within the framework of ACM."[23] **Table 1** identifies the 13 SIGs created in the 1960s (SIGMICRO, which was inactive for several of the middle decades, was not included).

---

[21] See Thomas J. Misa, ed., *Communities of Computing: Computer Science and Society in the ACM* (New York: ACM Books, 2016).

[22] Merlin G. Smith, "IEEE Computer Society: Four Decades of Service, 1951-1991," *Computer* 24 (September 1991): 6-12 at doi.org/10.1109/2.84894 .

[23] Jean Sammet, "Chapters and SIG/SICs," *Communications of the ACM* 19 no. 1 (Jan. 1976): quote 1 at doi.org/10.1145/359970.359973.

Table 1: ACM Special Interest Groups

| Special Interest Group | Founding/creation as SIG |
|---|---|
| SIGMIS (Management Information Systems) | 1961 |
| SIGUCCS (University [and College] Computing Centers) | 1963 [1981] |
| SIGIR (Information Retrieval) | 1963 |
| SIGOPS (Operating Systems) | 1965/1968 |
| SIGDA (Design Automation) | 1965/1969 |
| SIGART (Artificial Intelligence) *now SIGAI* | 1966 |
| SIGSAM (Symbolic and Algebraic Manipulation) | 1966 |
| SIGPLAN (Programming Languages) | 1967 |
| SIGGRAPH (Computer Graphics) | 1967/1969 |
| SIGCOMM (Data Communication) | 1967/1969 |
| SIGACT (Algorithms and Computation Theory) | 1968 |
| SIGCSE (Computer Science Education) | 1968 |
| SIGSIM (Simulation) | 1969 |

Source: historywiki.acm.org/sigs/Main_Page (and links to SIG pages)

The core data for this article are the research publications for each of these 13 SIGs in 1970, 1980, 1990 and 2000 accessed through the ACM's Digital Library. The year 2000 was pragmatically chosen as a terminal date for data collection; after that, the rising tide of SIG research publications simply becomes too large for my mixed-method of analysis. A comparative study of IEEE Computer Society publications might be attempted, but the IEEE CS did not create durable SIGs (an evolving set of sponsored "conferences" was historically and is today its chief subfield activity).

Existing research on gender bias in computing normally uses tabulations of women as a percentage of the authors (or members) in a given community. Accordingly, this established convention has significant merit in facilitating comparative analysis with existing datasets from NSF, CRA, the Census Bureau, and other sources. It sometimes can be helpful to examine the This is a pre-copyedited article published in *Information & Culture* 59 no. 2 (2024): 161-81. The publisher-authenticated version is available from University of Texas Press at DOI and Project Muse.actual number of women, since in a period of strong growth there may be increases in the total number of women gaining degrees, publishing articles, or getting jobs even if their proportion or percentage is not rising; yet percentages among groups and subgroups are more easily compared, and helpful for assessing changes in gender composition across time. Alternative measures, such as the percentage of all women graduates who gain degrees in the specific field of computer science may be instructive as a rough measure of the "attractiveness" of the field to women, but there are significant difficulties in determining the proper denominator for such a computation (all women? only US women? which colleges and universities?).

For each analyzed year, all research articles published by that SIG (and accessed through the ACM online Digital Library in July 2022) were collected for analysis, across all SIGs and years (N = 5,665). The ACM Digital Library provides a means to identify many "initials-only" authors, since ACM authors may add first/given names to their author's information, even years later; it also allows the reliable following of authors who might publish under slightly different names or who may change their names, so long at the author updates their biographical entry in ACM DL. (Not all authors wish such identification.[24]) The ACM DL also offers useful metadata, such as authors' institutional affiliations and collaborating authors, that helps accurately resolve individuals with common or similar names, again improving robust identification of thousands of authors.

The analysis of ACM SIG research publications proceeded as follows. All names from multiple-authored articles were extracted, and multiple instances of authorship in each year were combined to form the population of research authors (N = 7,456). Multiple authorship in computer science was expanding (from an average of 1.41 authors/article around 1960 to a range of 2.53 to 4.18, depending on subfield, after 2000), and conventions about "first" or "last" authorship do not appear to be stable.[25] The author data results in a measure of women as a percentage of the published authors for each analyzed year—and for each analyzed SIG. With care, the research authors' given/first names can be analyzed for gender (see note on method below). For the earliest articles around 1970, the ACM DL sometimes digitized only a portion of a SIG's annual research activity (conference papers and published articles); sometimes, too, scanning appeared to be incomplete. For these select instances, the set of articles was intentionally over-sampled (±1 or ±2 years), analyzed for authors' gender, then scaled back to the original author population size. Expanding the sample window (±1 or ±2 years) helped minimize the possibility of undue sample bias when there was incomplete data.

---

[24] Shelley K. Erickson, "Women Ph.D. Students in Engineering and a Nuanced Terrain: Avoiding and Revealing Gender," *Review of Higher Education* 35, no. 3 (Spring 2012): 355-374 at doi.org/10.1353/rhe.2012.0019 .

[25] João Fernandes and Miguel Monteiro, "Evolution in the Number of Authors of Computer Science Publications," *Scientometrics* 110 (2017): 1-11 at doi.org/10.1007/s11192-016-2214-9

Existing research with computer-science populations under or around 100 have sometimes utilized extensive personal look-ups of individual authors,[26] while studies with larger populations have typically used software to infer the genders of thousands of authors.[27] These gender-identification software packages must be used with caution, however, since their use for historical research can lead to errant findings and erroneous results. Powerful tools must be used with care.

Researchers in digital humanities have found that first names, which are key inputs for gender-identification software, do not have stable gender associations. Names change gender across time. Widely used software tools, with few exceptions, rely on present-day name–gender associations to "predict" the gender of historical names, leading to inaccurate and wayward results. Cameron Blevins and Lincoln Mullen aptly label the gender shifts evident in seven common given names during 1930–2012 as the "Leslie problem." For instance, the first name Leslie had been assigned at birth to 92% male babies in 1900, yet by 1950 it was used to name both male (48%) and female (52%) babies, while since 2000 it became a common female name (>96%).[28] With gender-prediction software relying on today's name–gender associations, Leslie Valiant and Leslie Lamport (male ACM Turing laureates in 2010 and 2013) are confidently but erroneously classified as female: 92% probability of being female according to Gender-API; 87% by Nam-Sor; and 77% by Genderize.io. By contrast, female computer-science luminaries such as Leslie Ann Goldberg and Leslie Pack Kaelbling (full professors at Oxford and MIT, respectively) would be more accurately gender identified. In a separate article, the author has expanded Blevins and Mullen's analysis to assess 13 million given US Social Security Administration (SSA) names, identifying the 50 top gender-changing names 1925–1975: the significant and measurable "net female" shift is consequential for software-based gender analysis of historical persons.[29]

---

[26] Laine Nooney, Kevin Driscoll, Kera Allen, "From Programming to Products: *Softalk* Magazine and the Rise of the Personal Computer User," *Information & Culture* 55, no. 2 (2020): 105-129, at doi.org/10.7560/IC55201 ; and R. Arvid Nelsen, "Race and Computing: The Problem of Sources, the Potential of Prosopography, and the Lesson of *Ebony* Magazine," *IEEE Annals of the History of Computing* 39 no. 1 (2017): 29-51 at doi.org/10.1109/MAHC.2016.11 .

[27] J. McGrath Cohoon, Sergey Nigai, and Joseph "Jofish" Kaye, "Gender and Computing Conference Papers," *Communications of the ACM* 54, no. 8 (August 2011): 72-80 at doi.org/10.1145/1978542.1978561 ; Sandra Mattauch et al., "A Bibliometric Approach for Detecting the Gender Gap in Computer Science," *Communications of the ACM* 63, no. 5 (May 2020): 76-78 at doi.org/10.1145/3376901 . Earlier publications of the present author developed a "mixed method" (part qualitative, part quantitative) for gender analysis of early computer-user groups, professional-society membership lists, and computer-science authors in the respected DBLP dataset: see "Gender Bias in Computing," in William Aspray, ed., *Historical Studies in Computing, Information, and Society* (Springer 2019), 115-136, at doi.org/10.1007/978-3-030-18955-6_6 ; "Dynamics of Gender Bias in Computing," *Communications of the ACM* 64 no. 6 (June 2021): 76-83, at doi.org/10.1145/3417517 ; "Gender Bias in Big Data Analysis," *Information and Culture* 57 no. 3 (2022): 283-306 at doi.org/10.7560/IC57303 . One review of gender-identifying software is Fariba Karimi, Claudia Wagner, Florian Lemmerich, Mohsen Jadidi, and Markus Strohmaier, "Inferring Gender from Names on the Web: A Comparative Evaluation of Gender Detection Methods," *WWW'16 Companion* (April 11-15, 2016) at dx.doi.org/10.1145/2872518.2889385 .

Such computer-driven undercounting of historical males with names such as Addison, Allison, Courtney, Kendall, Leslie, Madison, Morgan, Shelby, Sydney, and literally scores more has likely resulted in (improperly) inflated estimations of women in early computer-science publications and conference papers. Again, the net "female shift" noted above means that there is directionality to the bias in research based on these gender-identification software packages. The unusual results reported by Lucy Wang and colleagues in *Communications of the ACM* may owe

---

[28] Cameron Blevins and Lincoln Mullen, "Jane, John . . . Leslie? A Historical Method for Algorithmic Gender Prediction," *Digital Humanities Quarterly* 9, no. 3 (2015): not paginated, https://www.digitalhumanities.org/dhq/vol/9/3/000223/000223.html and archived at https://web.archive.org/web/20231003050616/https://www.digitalhumanities.org/dhq/vol/9/3/000223/000223.html. Also noting a "female shift" is Herbert Barry and Aylene S. Harper, "Feminization of Unisex Names from 1960 to 1990," *Names* 41 no. 4 (December 1993): 228-238 at doi.org/10.1179/nam.1993.41.4.228. A different software package to analyze (contemporary) first and last names to predict individuals' ethnicity is Fangzhou Xie, "Rethnicity: An R Package for Predicting Ethnicity from Names," *SoftwareX* 17 (2022): 100965 at doi.org/10.1016/j.softx.2021.100965 (thanks to Nabeel Siddiqui for this reference).

[29] Thomas J. Misa, "Temporal Analysis and Gender Bias in Computing," *arXiv* (October 2022) at doi.org/10.48550/arXiv.2210.08983

to their study's reliance on the Gender-API software tool.[30] Whereas previous studies have found women to constitute roughly 2% of computer-science research authors in the early 1950s,[31] Wang and colleagues offer a figure ten times higher. They report women to be 20% or more of computer-science authors. Their unusual findings for several other male-dominated research fields in the 1950s, such as engineering and medicine, may also stem from their use of gender-prediction software that was, in effect, "trained" on present-day name–gender associations but then became substantially mis-calibrated when applied to historical names. (Other issues may cloud their results.[32])

    The analysis developed for this article judiciously draws on an immense historical dataset from the US Social Security Administration, which annually lists since 1880 all given names (used five or more times each year) and tabulates their use for male and female babies. There are several reasons to approach the SSA dataset with some caution, but at least since the 1940s (when the SSA's earlier undercounting of female names as well as agricultural workers was corrected) it is simply the largest and most representative US national dataset available.[33] For each given/first name, the SSA tabulates the number of male babies and female babies who were given that name. For instance, Leslie Valiant was born in 1941 when there were 505 female babies and 1,557 male babies named Leslie, so one computes a probability of a random Leslie born that year being female, or p(F), as 0.24. This method of analysis, so long as an appropriate year of birth is chosen, offers results that are sensitive to historical changes in name–gender association. It does not properly recognize current non-binary gender identity practices. (Indeed,

---

[30] Lucy Lu Wang, Gabriel Stanovsky, Luca Weihs, and Oren Etzioni, "Gender Trends in Computer Science Authorship," *Communications of the ACM* 64, no. 3 (March 2021): 83 figure 4 at doi.org/10.1145/3430803.

[31] Women were roughly 1.8 percent of early CS authors (1949–54) according to José María Cavero et al., "The Evolution of Female Authorship in Computing Research," *Scientometrics* 103 (2015): 89 at doi.org/10.1007/s11192-014-1520-3. The author's findings (article cited above) from the DBLP dataset (1950-54) are 1.7 percent women. During 1966-70, women earned fewer than 3% of US doctoral degrees in computer science according to J. McGrath Cohoon and William Aspray, ed., *Women and Information Technology: Research on Underrepresentation* (Cambridge: MIT Press, 2006), p. x.

[32] Oddly, Wang et al. admit: "We also filter out first names that occur less than 10 times in our overall corpus, to reduce the number of API calls to manageable numbers." Why the Allen Institute for AI would trim such "edge cases" is unclear.

[33] For names *before* 1930, some researchers advocate use of the Integrated Public Use Microdata Series (IPUMS); from 1940s onward, many researchers and software packages, rely on the SSA data.

some readers may not be comfortable with my use of this gender-binary data; I would offer the suggestion that better understanding of gender bias in computing is a pressing public issue, and so the careful use of *some data* is needed.)

Since name–gender associations change across time, and since use of present-day associations would produce historical results that are in error, the analysis done for this dataset needs some "tuning" to yield acceptable results. What year is the appropriate one for the SSA lookup for an author publishing in, say, 1970? Clearly, using the year 1970 would be one option, but the author's year of birth (if only if it were known) would be ideal. How practically to proceed? In a previous article, I created a sizable dataset (N = 10,000) with computer-science authors drawn from the respected DBLP database (1950–80); extensive and time-consuming research resulted in personally identifying 80-100% of these authors. Their genders often could be found from publicly available biographical information, author-maintained websites, and other public reliable sources: Hilary Putnam, for example, was a well-known male logician. For this present article, four sub-groups from the DBLP paper (N = 101, 195, 137, and 114) with personally identified authors, and publicly known genders, were used as test datasets to calibrate the SSA lookup. For each test sub-group, I computed the gender for each author based on the SSA dataset—using seven different hypothetical "year shifts," from 20 to 50, prior to the article's year of publication. **Figure 2** indicates that three of the four testing scenarios identified 30 as the optimum "year shift" (minimizing the |absolute value| of each computed name's p(F) subtracted from the p(F) values of the personally identified name). A "year shift" of 30 is moreover a reasonable real-world parameter; few computer scientists begin publication prior to reaching 20 years of age, while many are active by age 30.

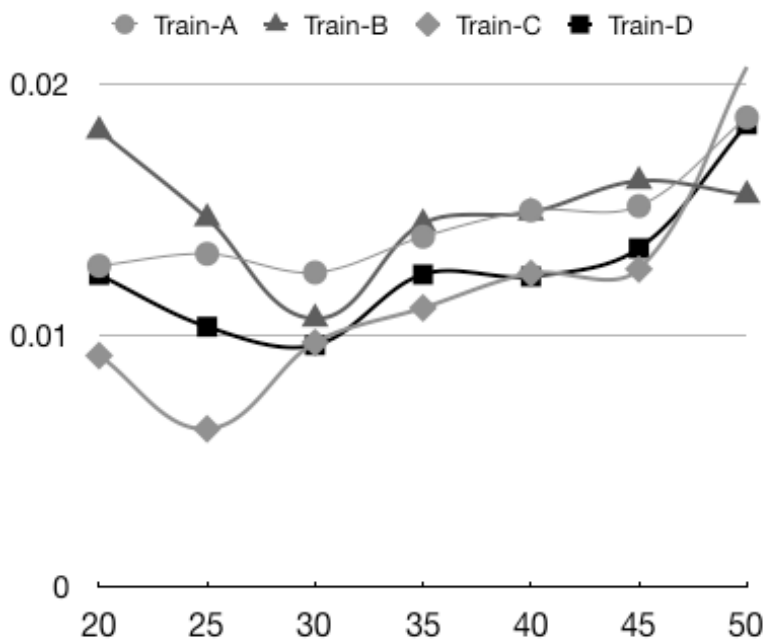

Figure 2: Training subgroups differential vs. "year-shift" for SSA lookup

Note: For subgroups A, B, and D, the minimum of | computed p(F) - actual p(F) | for tuning variable "year-shift" was 30 (i.e., SSA look-up year before year of publication). For subgroup C, it was 25. The article's analysis thus utilized the article's year of publication minus 30 years for the SSA lookup.

Another important consideration is reliably identifying genders for non-American, indeed non-Western authors. Already by 1980 ACM had numerous European authors with Finnish, German, Turkish, or Russian first names not appearing in the SSA data; and in later decades, increasing numbers of South Asian and East Asian-named authors.[34] One result of globalization is that numerous "non-SSA" names in 1970 had become common US names by 2000. While the increasing proportion of non-SSA names in 2000 (and, likely, beyond) means alternative means for gender analysis may be needed for further research, it is revealing that individual SIGs — once again — varied substantially in the percentage of "non-SSA" names used by their authors.

---

[34] A recent publication uses contextual Chinese-language insights to develop a method to predict genders for Chinese names, see Zihao Pan, Kai Peng, Shuai Ling, and Haipeng Zhang, "For the Underrepresented in Gender Bias Research: Chinese Name Gender Prediction with Heterogeneous Graph Attention Network," *Proceedings of the AAAI Conference on Artificial Intelligence* 37 no. 12 (2023): 14436-14443, at doi.org/10.1609/aaai.v37i12.26688

In 2000, for instance, SIGSIM had just 4.8% "non-SSA" authors and SIGUCCS 5.3%, while SIGGRAPH had 25.7%, SIGOPS, 27.7%, and SIGIR 34.9%. On this simple measure there are significant differences between the SIGs with lower percentages of "non-SSA" names, possibly a reflection of low or modest international participation, and those SIGs with higher "non-SSA" names and likely greater international participation.

Clearly, the US-centered SSA dataset of name–gender associations is not applicable to authors from beyond the US. To partly correct this source of error, the analysis incorporated contextual factors, using internet searching to personally verify individuals with names such as Andrea, Jan, Jean, Joan, Laurence and numerous others that vary culturally in gender associations (e.g. Jean and Andrea were common women's names in the US but more frequently men's names in Francophone or Italian countries). Generally, the method of analysis reversed findings from the SSA lookups only when there was positive personal identification, such as Andrea Asperti at the University of Bologna; there were plenty of North American female Andrea's as well.

**Dramatic Differences in Women's Authorship**

The ACM data reveals dramatic differences in women's changing participation in its SIG publications, both across time and between SIGs. The principal empirical findings can be briefly summarized: [a] women's participation as research authors during 1970-2000 increased in all SIGs; [b] individual SIGs, however, varied markedly in their levels of women's participation (ranging from 0% upwards to nearly 50%) as shown in **Figures 3 and 4**; [c] most but not all SIGs experienced decelerating growth in women's participation; [d] individual SIGs can be benchmarked against a composite ACM, in their median percentage of women authors and in the shape of their growth curves (convex, linear, or concave as reflecting, respectively, accelerating, linear, or decelerating growth) as shown below in **Figure 5**.

Figures 3 and 4, depicting temporal changes, present three variables. The x-axis is the analyzed years 1970 to 2000; the y-axis is the computed percentage of women as research authors, based on this article's analysis of SSA data with several contextual refinements (as noted above and discussed below), and the bubble area represents the number of all SIG authors (male, female, and unidentified). The 7 smaller SIGs appear in Figure 3 while the 6 larger SIGs, with their substantially larger populations, appear in Figure 4. Literally all SIGs started in 1970 with small women's research-article authorship, no more than 10% with several effectively at 0%. Most every SIG grew in size and all of them expanded women's research-article authorship, more or less monotonically, across these four decades. ACM experienced substantial growth in women's research authorship across 1970, 1980, 1990, and 2000: respectively, 3.6%, 12.4%, 15.1%, and 17.3% (figures are composites of the 13 SIGs).

**[Figure 3 here]**

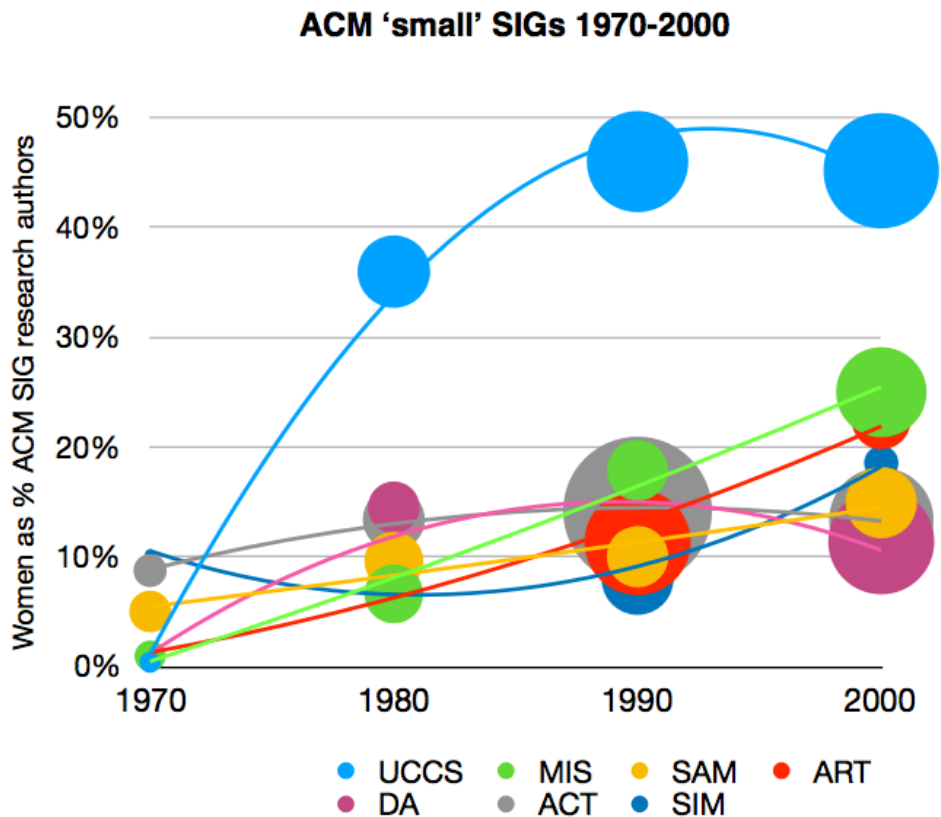

[Figure 4 here]

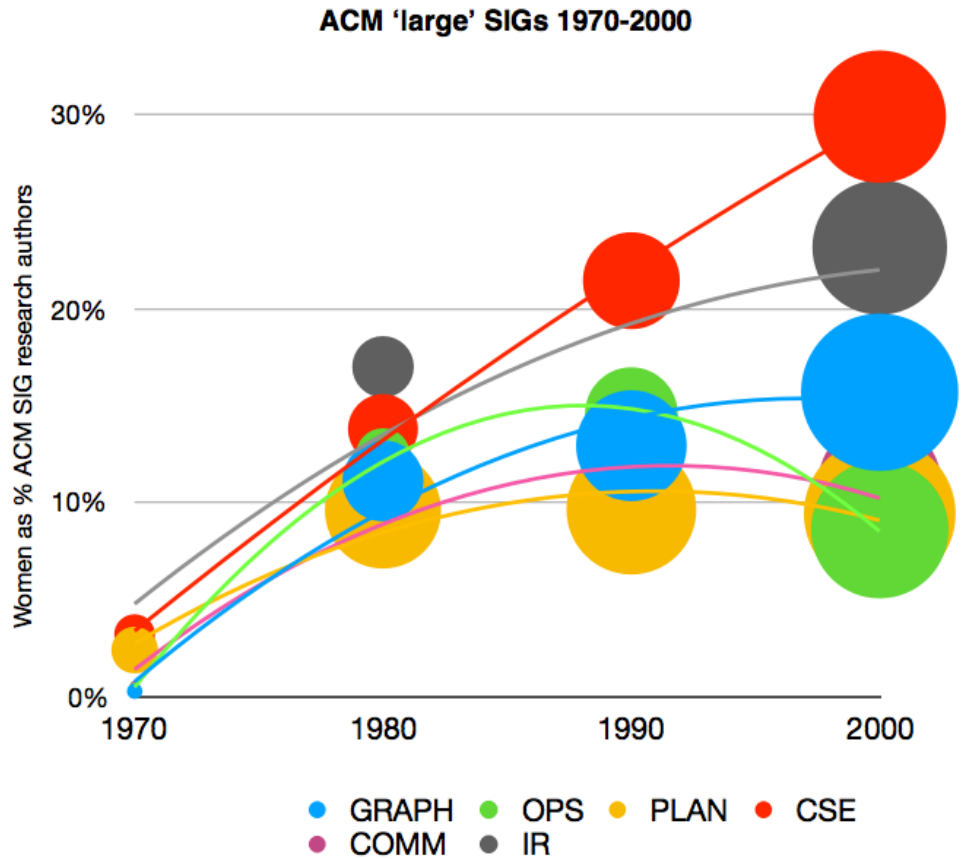

Analysis of trendlines for the growth curves indicated that the data for SIG research articles are well described using second-order polynomial trendlines (with the median for computed $R^2$ values for "goodness of fit" between actual data points and trendline curves at a highly respectable 0.97). Alternate trendlines using linear growth ($R^2$ median value 0.78) were less robust, reflecting that many SIGs, notably SIGSIM, SIGDA, and SIGOPS (respectively 0.41, 0.40, 0.31 for $R^2$ values), simply did not experience "linear" growth. All SIGs, as well as the composite ACM, grew in women's research authorship; but they did not grow equally.

The second-order polynomial trendlines offer a measure of the "shape" of the growth curves, indicating the accelerating or decelerating levels of authorship of women in each of the SIGs. Generally, the "shape" of curves gives rise to a complex body of mathematics used in financial analysis, economic theory, and even biology. These curves for women's authorship in ACM SIGs are comparatively straightforward, indeed nearly monotonic; the coefficients for the trendline's $x^2$ polynomial term are a ready measure of the curves' accelerating growth, linear growth, or decelerating growth; in other words, of the curve's convexity, linearity, or concavity (see **Figure 5**). Accelerating growth (convex trendlines for women's authorship) was rare, found only in SIGSIM, SIGART, and SIGMIS; these three SIGs were among the lowest in median women's authorship (each below 10%).

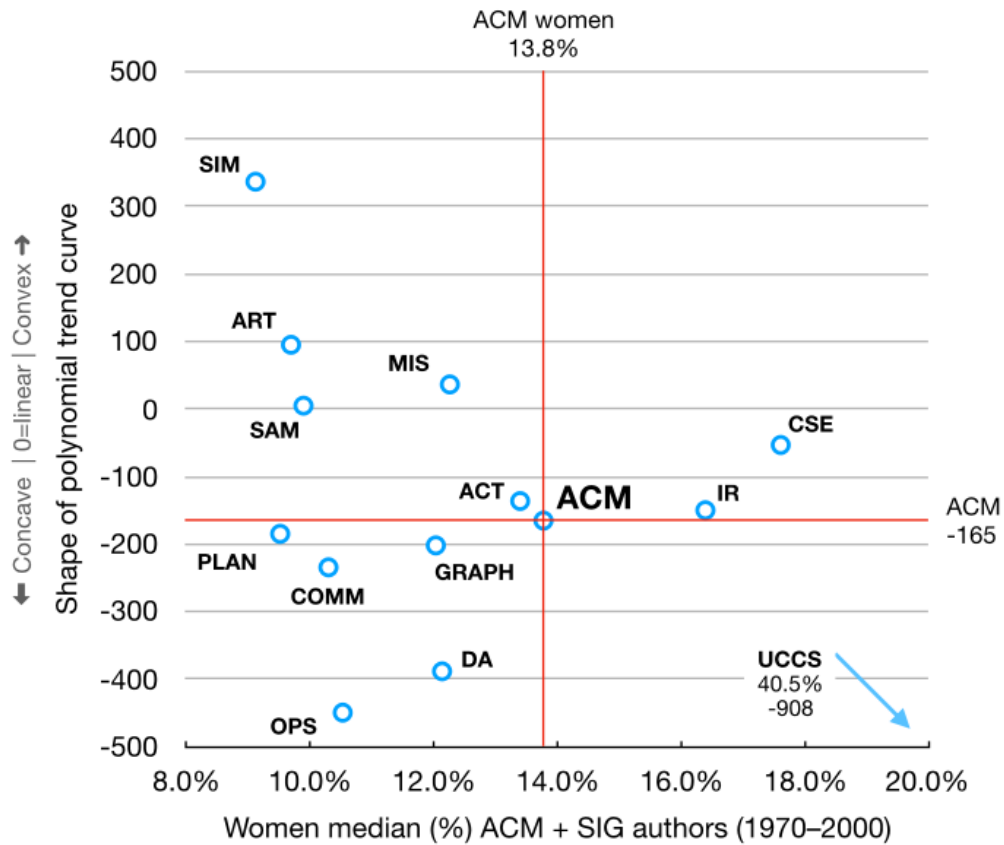

Figure 5: Median % of women authors (1970-2000) vs. 'shape' of growth curve

A means to benchmark each SIG against a composite ACM is reasonably simple. Median figures for women's participation in each SIG across 1970–2000 are easily computed; and a weighed average for a composite ACM is then close at hand. The composite ACM had a median women's authorship of 13.8% and a markedly decelerating or concave growth curve (-165). This composite ACM can be placed at the center of a quadrant (see red lines in **Figure 5**) by which the 13 SIGs may be compared.

One outlier needs to be mentioned: SIGUCCS (university and college computing centers) was literally off the chart. While other SIGs clustered around 10-18% women's authorship, SIGUCCS's median across 1970-2000 was 40.5%. While it started with virtually no women in 1970, it grew quickly in both size and in women's authorship (see Figure 3), peaking at 45.9% women authors in 1990 and maintaining 45.1% in 2000. Figure 5 might at first appear odd, since the ACM composite women's authorship looks to be anomalously high, with most SIGs appearing to be lower, a mathematical impossibility. But the ACM composite at 13.8% properly incorporates the off-the-chart high SIGUCCS data; recomputing the ACM composite by removing SIGUCCS results in a (entirely hypothetical) median fully two percentage points lower at 11.8%, square in the middle of the other plotted SIGs.

**Conclusion**

It may not be comfortable to cast this data on women's authorship in computer science as "gender bias," but with such sharply varying rates of women's authorship — painfully modest to nearly 50% — it is clear that ACM SIGs varied dramatically in their levels of women as research-article authors. Some SIGs, especially UCCS, but also SIGIR and SIGCSE (16.4% and 17.6% respectively) had above-median women's authorship and so seeming evinced openness to women's participation in the research community. Conversely, other SIGs were significantly below the ACM median in women's authorship including SIGSIM, SIGART, SIGPLAN, and SIGSAM (respectively, 9.1%, 9.7%, 9.5%, and 9.9%). Additional research in SIG historical files is needed to assess any SIG's organizational culture, its policies and procedures, and its possible barriers to and/or encouragement of women's participation and research authorship.[35]

This dataset demonstrates that "computer science" is nothing like a unitary field, at least with respect to changing levels of women's participation in the research community. Instead of positing changes that "computer science" might make as a whole (a common strategy for reform efforts), it may be better to recognize that the CS subfields represented by the ACM SIGs have different histories, distinct cultures, and divergent experiences with gender dynamics. The ACM subfields are not the same, and require consideration of their specific cultures. Given this data, it is fruitful to contrast the six larger SIGs with the seven smaller SIGs; computer science grew strongly during these decades, and the gender dynamics of the six large SIGs — with fully 75 percent of the research authors examined here — certainly had an outsize influence on the expansion and the gender dynamics for ACM as a whole.

It is sometimes suggested that male-dominated fields such as computer science in effect serve to select men positively and, accordingly, to discourage women. Understandably, women may prefer to avoid a gender-slanted educational or work environment; while, possibly, some men might prefer such an environment, a factor that may sustain "bro" culture.[36] To make a solid judgment, one needs to consult archival materials about individual SIG's membership, officers, and internal operations (this archival material is practically nonexistent at present), but the comparative data presented here are suggestive. Four of the large SIGs had women's authorship that was below the ACM median of 13.8% as well as decelerating growth (concave growth curves): these are PLAN, COMM, GRAPH, and OPS (respectively, programming languages, communications, graphics, and operating systems), which together account for 49% of the sampled research authors. Of the large SIGs only CSE was substantially above the ACM medians for growth and women's research authorship. Similarly, the two ACM SIGs with the largest median women research authors were CSE (17.6%) and UCCS (40.5), each of which had positive reasons to attract women: CS education and university and college computing centers, respectively. These two SIGs, accounting for just 20% of SIG research authors, had 32% of the women research authors analyzed in this study.

---

[35] The ACM's History Committee supports SIGs in identifying and archiving their records; presently, the needed archival records to evaluate SIG's changing membership, officers, and internal procedures simply do not exist. See "2019 ACM SIG Heritage Workshop Held at Charles Babbage Institute" (17 July 2019) https://history.acm.org/2020/02/01/2019-acm-sig-heritage-workshop-held-at-charles-babbage-institute/ and "Capturing Hidden ACM History" (30 September - 1 October 2022) at https://history.acm.org/seminar-capturing-hidden-acm-history/ (accessed Feb. 2023)

[36] Emily Chang, *Brotopia: Breaking Up the Boys' Club of Silicon Valley* (New York: Portfolio/Penguin, 2019) ; Erin Griffith, "Silicon Valley Slides Back Into 'Bro' Culture," *New York Times* (24 September 2022) archived at https://web.archive.org/web/20230905100826/https://www.nytimes.com/2022/09/24/technology/silicon-valley-slides-back-into-bro-culture.html

Put another way, for ACM SIGs with above-median women's authorship (CSE, UCCS, and IR), the "reform efforts" could best be focused on *retention* of women already present as research authors and members of the subfield community; conversely, for SIGs with below-median women's authorship (the four lowest are SIM, ART, PLAN, and SAM, respectively, simulation, artificial intelligence, programming languages, and symbolic and algebraic manipulation), reform could instead focus on *recruitment* of women and possible culture change, conceivably along the lines of Carnegie Mellon's "cultural approach" to increasing women in computer science. A more nuanced approach to the gender diversities within "computer science," such as advocated here, would recognize the evident differences in gender composition across the subfields of computer science, here represented by ACM SIG research publications and authors.

This research might be extended in four different and complementary directions. First, the SIGs that were organized after 1970, such as SIGCAS, SIGMOD, SIGMETRICS (respectively, computers and society, management of data, and computer performance evaluation), and others might be analyzed using this same mode of analysis. Second, these and other SIGs might be analyzed in years beyond 2000, although the enormous numbers of research articles and research authors make this a frankly daunting prospect. Third, further research is needed to better understand emerging CS subfields that developed alongside and, sometimes, independently of ACM, evident in seemingly anomalous decreases (after c. 1990) in the total research authors for some ACM SIGs.[37] Finally, for deeper insight into the dynamics of gender bias, future research into the ACM SIGs' organizational records, conference proceedings, and oral histories of officers and members should lead to a better understanding of what some SIGs did right in facilitating women as research authors and members of the computer science community.[38] Those lessons are certainly needed today.

___

**Author Declaration:** The author declares no conflicts of interest.

**Acknowledgements:** This line of research began at the Charles Babbage Institute, where my colleague Jeff Yost (now CBI director) pointed out to me that one of the leading computer user groups (SHARE, founded 1955), had extensive archival lists of its early members and attendees, across multiple years, with individual members listed by their *first* names.

___

[37] For example, SIGSAM in 1989 spawned the International Symposium on Symbolic and Algebraic Computation (ISSAC), which (in 2023) is still "sponsored" by ACM SIGSAM with full acknowledgement of its founding within ACM. By comparison, SIGART (renamed SIGAI in 2014) cooperates at greater distance with the independent Association for the Advancement of Artificial Intelligence (AAAI), founded in 1979 as the American Association for Artificial Intelligence; see Yolanda Gil, "SIGART to SIGAI," SIGAI Newsletter (January 2014) archived at https://web.archive.org/web/20140128074304/http://sigai.acm.org/activities/misc/sigai_appointed.html ; Sven Koenig, Sanmay Das, Rosemary Paradis, John Dickerson, Yolanda Gil, Katherine Guo, Benjamin Kuipers, Iolanda Leite, Hang Ma, Nicholas Mattei, Amy McGovern, Larry Medsker, Todd Neller, Marion Neumann, Plamen Petrov, Michael Rovatsos, and David Stork, "ACM SIGAI activity report," *AI Matters* 5 no. 3 (September 2019), 6-11, at doi.org/10.1145/3362077.3362079 . A different source of year-by-year variance is that SIG-organized conferences sometimes were indexed completely, article by article, in the ACM DL; other times, there is only one entry in the ACM DL, for the entire conference itself, consequently skewing the apparent number of conference papers and research authors. One key question may be: did the SIGs, embedded in ACM's organizational culture, have (for instance) greater, lesser, or roughly the same levels of women's participation as the "independent" conferences that they spawned or even competed with?

[38] With the Computing Educators Oral History Project (CEOHP), SIGCSE members have created a valuable set of three dozen oral histories: see Vicki L. Almstrum, Barbara Boucher Owens, Mary Z. Last, and Deepa Muralidhar, "CEOHP Evaluation, Evolution, and Archival Storage," in *Proceedings of the 43rd ACM technical symposium on Computer Science Education* (SIGCSE '12) (New York: Association for Computing Machinery, 2012), 674 at doi.org/10.1145/2157136.2157405 ; the CEOHP interviewees are listed, with biographical information and PDFs of the interview, at https://ceohp.heritage.acm.org/ceohp-collection/ and archived at https://web.archive.org/web/20230923014610/https://ceohp.heritage.acm.org/ceohp-collection/.